\pdfoutput=1
\documentclass[10pt,journal,compsoc]{IEEEtran}

\usepackage{cite}
\usepackage{amsmath,amssymb,amsfonts}
\usepackage{algorithmic}
\usepackage{graphicx}
\usepackage{textcomp}

\usepackage{booktabs}
\usepackage{threeparttable}
\usepackage{multirow}
\usepackage{bigstrut}
\usepackage{subfigure}
\usepackage{array}
\usepackage{algorithm}
\usepackage{enumerate}
\usepackage{colortbl}
\usepackage{float}
\usepackage{fancyhdr}
\usepackage{threeparttable}
\usepackage{ragged2e}
\usepackage{url}
\usepackage[table]{xcolor}
\usepackage{stfloats}
\usepackage{setspace}
\usepackage{fancyhdr}
\usepackage{etoolbox}
\usepackage{diagbox}
\usepackage{caption}
\usepackage{float}
\usepackage{booktabs}
\usepackage{footnote}

\newcommand{\tabincell}[2]{\begin{tabular}{@{}#1@{}}#2\end{tabular}}

\begin{document}

\title{Use the Spear as a Shield: A Novel Adversarial Example based Privacy-Preserving Technique against Membership Inference Attacks}

\author{Mingfu~Xue,
        Chengxiang Yuan,
        Can He,
        Zhiyu Wu,
        Yushu Zhang,
        Zhe Liu,
        and Weiqiang~Liu%

\IEEEcompsocitemizethanks{
\IEEEcompsocthanksitem {M. Xue, C. Yuan, C. He, Y. Zhang and Z. Liu are with the College of Computer Science and Technology, Nanjing University of Aeronautics and Astronautics, Nanjing, 211106, China. \protect\\
E-mail: \{mingfu.xue, yuancx, hecan, yushu, zhe.liu\}@nuaa.edu.cn.}

\IEEEcompsocthanksitem{ Z. Wu is with the College of Science, Nanjing University of Aeronautics and Astronautics, Nanjing, 211106, China. \protect\\
E-mail: wuzhiyu@nuaa.edu.cn.}

\IEEEcompsocthanksitem{ W. Liu is with the College of Electronic and Information Engineering, Nanjing University of Aeronautics and Astronautics, Nanjing, 211106, China. \protect\\
E-mail: liuweiqiang@nuaa.edu.cn.}
}
}

\IEEEtitleabstractindextext{
\begin{abstract}
\justifying
Recent researches demonstrate that machine learning models are vulnerable to privacy leakage attacks. Among them, the membership inference attack poses a serious threat to the privacy of confidential training data. In the membership inference attack, the adversary uses a membership inference model to determine whether a given data belongs to the training set of the target model based on the prediction of the target model. Few defenses have been proposed, but suffer from compromising the performance or quality of the target model, or cannot effectively resist against membership inference attacks.
This paper proposes a novel adversarial example based privacy-preserving technique (AEPPT), which adds the crafted adversarial perturbations to the prediction of the target model to mislead the adversary's membership inference model. The added adversarial perturbations do not affect the accuracy of target model, but can prevent the adversary from inferring whether a specific data is in the training set of the target model.
Since AEPPT only modifies the original output of the target model, the proposed method is general and does not require modifying or retraining the target model.
Experimental results show that the proposed method can reduce the inference accuracy and precision of the membership inference model to $50\%$, which is close to a random guess. The recall of the membership inference model drops from $88.7\%$ to $51.9\%$ on the CIFAR100 dataset and drops from $98.5\%$ to $17.1\%$ on Purchase dataset.
Moreover, the performances of the proposed method under various factors (\textit{i.e.}, perturbation step size, number of adversary's data, number of target model's output classes, and different membership inference models) are evaluated, which demonstrate that the proposed method can resist membership inference attacks under different factors.
Further, for those adaptive attacks where the adversary knows the defense mechanism, the proposed AEPPT is also demonstrated to be effective.
Compared with the state-of-the-art defense methods, \textit{e.g.}, the truncating method, the dropout method, and the adversarial regularization method, the proposed defense can significantly degrade the accuracy and precision of membership inference attacks to 50\% (\textit{i.e.}, the same as a random guess). Meantime, the normal performance and utility of the target model will not be affected.
\end{abstract}

}

\maketitle
\IEEEpeerreviewmaketitle

\section{Introduction}\label{sec:introduction}

\IEEEPARstart{M}ACHINE learning techniques are widely used in many fields, such as image classification, natural language processing and financial analysis, \textit{etc}.
As an increasingly popular business model, many companies (\textit{e.g.}, Google and Amazon) deploy machine learning as a service (MLaaS) for various clients, such as data processing, model training and data prediction. Users can upload data to these service providers to construct machine learning models, or use these models through prediction application programming interfaces (APIs).

However, recent studies \cite{shokri2017membership,salem2018ml,nasr2018machine,long2018understanding, yeom2018privacy} have shown that attackers can infer the training data information of the machine learning model through membership inference attacks \cite{shokri2017membership}.
The attackers first train a membership inference model, which is essentially a binary classifier.
Then, according to the output of the target model, attackers use the trained membership inference model to determine whether a given data belongs to the training set of the target model \cite{shokri2017membership}. If an adversary can correctly infer that a specific data has been used to train the target model, the target model has the risk of privacy leakage \cite{shokri2017membership}.

To date, only a few defense techniques against membership inference attacks have been proposed, which can be divided into two categories. The first method is to modify the output of the target model so that the model leaks less private information, such as the rounding method \cite{shokri2017membership} (also called as the truncating method). The second method is to prevent the target model from overfitting, such as the dropout method \cite{salem2018ml}, model stacking \cite{salem2018ml}, $L_{2}$-norm regularization \cite{shokri2017membership}, and adversarial regularization \cite{nasr2018machine}. However, the first method can only slightly mitigate membership inference attacks, and will affect the utility of predictions (because only a very rough confidence information are provided to the user).
For example, if the confidence score of two labels in a prediction is 0.2501 and 0.2599, the prediction of truncating method (with 2 decimals) on these two labels will be the same (\textit{i.e.}, 0.25), which makes the final prediction inaccurate.
The second method needs to modify the target model or modify its training process, and requires to sacrifice the model's performance to resist against membership inference attacks.

In this paper, we propose a novel \underline{a}dversarial \underline{e}xample based \underline{p}rivacy-\underline{p}reserving \underline{t}echnique (AEPPT) against membership inference attacks.
Inspired by adversarial examples that can mislead machine learning models, we convert the prediction of a target model into the adversarial prediction to mislead attackers' membership inference models.
First, the AEPPT trains a substitute membership inference model using member data (that has been used to train the target model) and non-member data (that has not been used to train the target model).
Then, according to the output of the substitute membership inference model, AEPPT crafts small perturbations that can mislead the attacker but do not affect the utility of the predictions.
In our experiments, we exploit the $L_1$-norm \cite{carlini2017adversarial} to calculate the size of the added perturbations. The smaller the value of the $L_1$-norm is, the smaller the adversarial perturbations are added, \textit{i.e.}, the difference between the adversarial prediction and the original prediction is more difficult to be noticed.
Finally, AEPPT adds the perturbations to the prediction of the target model to generate adversarial prediction, which makes the attacker unable to correctly determine whether a given data is in the training set of the target model.

The major contributions of this paper are as follows:
\begin{enumerate}
  \item{\textbf{A novel privacy-preserving technique that uses adversarial examples to resist membership inference attacks is proposed.}
      Adversarial example used to be an attack method targeting machine learning models. However, in this work, we are doing the opposite.
      We take advantage of adversarial examples to protect the privacy of the training data of the model.
      The proposed method adds perturbations on the original prediction of the target model to construct the adversarial prediction, which can successfully mislead the attacker's membership inference model.
      Experimental results show that the proposed method can make the inference accuracy of three different membership inference models reduce to a random guess (\textit{i.e.}, $50\%$). It also greatly reduce the precision and recall of different membership inference models of the attackers.}

  \item {\textbf{The proposed method does not modify or retrain the target model, thus is a general method and can be used to protect most machine learning models against membership inference attacks.}
  The proposed method only modifies the output of the target model to resist membership inference attacks without affecting the target model.
  In the experiment, we use the AlexNet network and the fully connected neural network to train the target model respectively, and then evaluate the proposed method on the two target models. For these two target models under the protection of the proposed method, the inference accuracy of adversary's membership inference attacks is only around $52\%$ and $51\%$ respectively, which are close to that of a random guess. Moreover, the accuracy of the target model with defense is close to the accuracy of the target model without defense.}

  \item {\textbf{The proposed method does not affect the utility of the prediction.}
  In existing works, the methods of modifying the output of the target model will make the final prediction inaccurate, \textit{e.g.}, truncating the confidence scores in the prediction to 2 or 3 decimals. Hence, these methods seriously affect the utility of the prediction, and reduce the quality of services \cite{lee2018defending}.}
  The proposed AEPPT only adds small size of adversarial perturbations (0.2637 in CIFAR100 \cite{krizhevsky2009learning} and 0.0868 in Purchase \cite{kaggle_challenge}) on the original prediction, which means that the average modifications on the confidence score of each class label is as low as 0.002637 (CIFAR100) and 0.000868 (Purchase), respectively.
  Therefore, the added perturbations will not affect the prediction of the target model, and the generated adversarial prediction is close to the original one. Meanwhile, the generated adversarial prediction can make the membership inference attack fail.

  \item{\textbf{The proposed method is robust under various factors and strong adaptive attacks.}
  First, the performance of the proposed method under different factors are evaluated, including: the size of added perturbations, the number of target model's output classes, the number of the adversary's data, and different membership inference models. Experimental results show that these factors will not affect the defensive performance of the proposed method, and the proposed AEPPT can resist membership inference attacks under different factors. Second, for those more complex adaptive attacks (flip attack, rounding attack \cite{shokri2017membership, jia2019memguard}, and adversarial training attack \cite{Athalye2018Obfuscated}), the proposed AEPPT method is also demonstrated to be robust and effective.}

\end{enumerate}

This paper is organized as follows.
Section \ref{related_work} reviews membership inference attacks and related defense works.
Section \ref{sec:Preliminary} describes the background of adversarial examples and the evaluation metrics used in this work.
Section \ref{sec:The_proposed_AEPPT} elaborates the proposed defense method. Experimental results are presented in Section \ref{sec:Experiments}. Section \ref{sec:conclusion} concludes this paper.

\section{Related Works}\label{related_work}
Earlier membership inference attacks were targeting at biomedical data or genomic data. For instance, some researches \cite{homer2008resolving, sankararaman2009genomic} have shown that a specific individual can be identified or inferred from statistics of the genome-wide association studies (GWAS).
Recently, membership inference attacks are applied to machine learning models. Shokri \textit{et al.} \cite{shokri2017membership} present the first membership inference attack in the context of machine learning models. They use the target model's prediction of member data and non-member data to train a membership inference model which acts as a binary classifier \cite{shokri2017membership}. The generated membership inference model can infer whether or not a specific data was used to train the target model. However, the attack method in \cite{shokri2017membership} makes a lot of assumptions, such as the adversary needs to know the structure and the training algorithm of the target model, and the data obtained by the attacker has the same distribution of the target model's training data. Hence, Salem \textit{et al.} \cite{salem2018ml} propose attack method with fewer assumptions, in which the attacker first uses a number of different models to mimic the behavior of the target model. Then, the attacker combines different model's predictions to perform membership inference attacks without the knowledge of the target model's structure and the training algorithm \cite{salem2018ml}. Hayes \textit{et al.} \cite{hayes2019logan} show that Generative Adversarial Networks (GANs) are also vulnerable to membership inference attacks. They train a GAN model to learn the information of the target generative model and use the trained GAN model to distinguish member data or non-member data of the target model's training set \cite{hayes2019logan}. However, if the discriminator in the GAN does not provide the query function, this method cannot attack successfully.

The defense against membership inference attacks can be divided into two categories: 1) change the confidence score in the prediction of the target model to a rough value; 2) prevent the target model from overfitting.

\textbf{Change the Confidence Score in the Prediction of the Target Model to a Rough Value.} Since the adversary performs membership inference attacks based on the prediction of the target model, Shokri \textit{et al.} \cite{shokri2017membership} mitigate membership inference attacks by modifying the prediction of the target model in the following ways. The first method is that the prediction returned by the target model only contains the confidence scores of the top $k$ classes \cite{shokri2017membership}. The second method rounds (\textit{i.e.}, truncates) the confidence score in the prediction and returns an approximate prediction confidence \cite{shokri2017membership}. The third method modifies the output layer of the neural network to reduce the correlation between the output and the input of the model \cite{shokri2017membership}.
Shokri \textit{et al.} \cite{shokri2017membership} demonstrate that these methods can slightly reduce the inference accuracy of the membership inference model, but cannot reduce the inference accuracy to 50\% (\textit{i.e.}, as a random guess). In other words, these methods cannot defend against membership inference attacks effectively.
In addition, these methods returns an approximate prediction to the users, which will affect the utility of the prediction and reduce the quality of the MLaaS \cite{lee2018defending}.
Compared with these methods proposed in \cite{shokri2017membership},
the proposed AEPPT only adds small adversarial perturbations to the predictions of the target model, which will not affect the utility of the prediction. Moreover, the added adversarial perturbations can significantly reduce the inference accuracy of the membership inference model to that of a random guess, \textit{i.e.}, 50\%, which means the attack completely failed.

\textbf{Prevent the Target Model from Overfitting.}
Some studies \cite{long2018understanding, yeom2018privacy} have shown that overfitting is a sufficient but unnecessary condition for the success of membership inference attacks.
In other words, overfitting is an important but not the only reason for the success of membership inference attacks \cite{shokri2017membership}.
Several existing methods resist membership inference attacks by preventing the target model from overfitting. Shokri et al.\cite{shokri2017membership} use $L_2$-norm regularization to mitigate the overfitting of the target model.
Salem \textit{et al.} \cite{salem2018ml} use dropout and model stacking to reduce the risk of privacy leakage of the target model's training data.
Specifically, the dropout method \cite{salem2018ml} drops a neuron with a certain probability in each training iteration.
The model stacking method \cite{salem2018ml} combines the output of multiple different models to resist against membership inference attacks.
However, the dropout method is only suitable for neural networks, and the model stacking method requires the defender to train multiple different models.
Nasr \textit{et al.} \cite{nasr2018machine} use adversarial regularization method to train the target model, which minimizes the classification loss of the model and the successful rate of membership inference attacks.
The defense mechanism of our proposed AEPPT is completely different from the adversarial regularization method \cite{nasr2018machine}.
First, the adversarial regularization method \cite{nasr2018machine} adds the adversarial regularization term to the loss function of the target model, which aims to defeat the membership inference attack by mitigating the overfitting of a target model.
The proposed AEPPT is inspired by the adversarial examples, and generates the adversarial prediction based on the target model's output, so as to mislead the adversary's membership inference model to make incorrect output.
Second, the adversarial regularization method \cite{nasr2018machine} needs to modify the training process of the target model, and their experiment results show that this will cause a drop in the prediction accuracy of the target model.
In other words, the work \cite{nasr2018machine} requires a trade-off between the performance of the target model and the ability to resist membership inference attacks.
However, the proposed AEPPT method only adds adversarial perturbations on the original output of the target model, thus it can successfully defeat the membership inference attack without affecting the model's performance.

\textbf{The Concurrent Work.}
We proposed the idea of this paper in May 2019, and applied for a Chinese patent \cite{Wu2019AImodel} in July 2019.
To date, the only concurrent work with this paper is MemGuard \cite{jia2019memguard}. Jia \textit{et al.} \cite{jia2019memguard} turn the prediction of the target model into adversarial examples to resist against membership inference attacks.
The method formulates the defense of membership inference attacks as solving the optimization problem. Specifically, they first search for noise that does not affect the utility of the target model but can mislead the membership inference models.
Then, they add the noise to the prediction of the target model with a certain probability \cite{jia2019memguard}.

There are several differences between the proposed method and the MemGuard method \cite{jia2019memguard}.
First, the MemGuard method generates adversarial perturbations by solving the optimization problem, \textit{i.e.}, the optimization-based method. The proposed method crafts adversarial perturbations by calculating the gradient of the loss function, \textit{i.e.}, the gradient-based method. Since solving the optimization problem requires expensive computational overhead, the proposed method generates adversarial predictions faster than MemGuard.
Second, researches have shown that the adversarial examples generated by the gradient-based method have better transferability than the adversarial predictions generated by the optimization method, which means that the proposed method can defend more membership inference models than MemGuard \cite{jia2019memguard}.
Third, the MemGuard method adds adversarial perturbations to the prediction of the target model with a certain probability.
The proposed AEPPT multiplies the perturbations by a random step size and adds it to the prediction of the target model.
Therefore, compared with the MemGuard method, the perturbations added by the proposed method are more difficult to be found by the adversary.
Lastly, compared with \cite{jia2019memguard} that assumes the attacker has only black-box access to the target model, the attacker of this paper has more knowledge and has stronger capabilities, which demonstrates that the proposed method can defeat more powerful attackers than work \cite{jia2019memguard}.
In conclusion, both the proposed AEPPT and the MemGuard \cite{jia2019memguard} can effectively resist against membership inference attacks, but the proposed AEPPT can resist more powerful attackers, and is more secure, more general and can generate adversarial predictions faster than the MemGuard method \cite{jia2019memguard}.
In Section \ref{sec:discussion}, we will compare and analyze the defense performance of our proposed AEPPT method and the MemGuard method \cite{jia2019memguard}.

\section{Preliminary}\label{sec:Preliminary}
In this section, we first describe the background of adversarial examples and the transferability of adversarial examples in Section \ref{sec:AE}. Then, the evaluation metrics used in this work are described in Section \ref{sec:Metrics}.

\subsection{Adversarial Examples}\label{sec:AE}
Ideally, given a trained machine learning model and an input, the model will output a predicted label. The adversarial example attack is to add small perturbations to the input to generate an adversarial example, which can cause the model to produce an erroneous output \cite{szegedy2013intriguing}.
Generally, adversarial example attacks can be divided into white-box attacks, grey-box attacks and black-box attacks. If the adversary has full knowledge of the target model, the adversarial example attack is a white-box attack \cite{akhtar2018threat}. Conversely, if the adversary does not have any knowledge of the target model, it is a black-box attack \cite{akhtar2018threat}. If the adversary has some knowledge of the target model (\textit{e.g.} knows the architecture information of the target model, but does not know the parameter settings of the target model), it is a grey-box attack \cite{zhang2019adversarial}.

\subsection{Evaluation Metrics}\label{sec:Metrics}

To demonstrate the defensive performance of the proposed method, we evaluate the performance of the membership inference model (from the attacker's perspective) and the target model (from the defender's perspective) with and without defense, respectively.
The similarity between the original prediction and the adversarial prediction is also evaluated. The evaluation metrics are as follows.

\textbf{Metrics for Evaluating the Membership Inference Model:}
\textit{Inference accuracy}, \textit{precision}, and \textit{recall} are used to evaluate the impact of the proposed method on the attacker's membership inference model.
Inference accuracy is the proportion of data (including member data and non-member data) that correctly inferred by the attacker among all the data \cite{shalev2014understanding}.
Precision is the proportion of the true member data among all inferred member data \cite{shalev2014understanding}.
Recall is the proportion of correctly inferred member data among all the member data \cite{shalev2014understanding}.

Similar to the experimental settings in \cite{shokri2017membership,nasr2018machine,song2019membership}, the number of member data and non-member data used to evaluate the membership inference attacks are set to be equal in this experiment. Hence, the inference accuracy and precision of an adversary using random guess is $50\%$, which is the lower bound of the inference accuracy and precision of an adversary.
For the recall of the adversary, the lower bound is $0\%$.
The goal of the proposed method is to make the inference accuracy and precision of the membership inference model close to that of a random guess ($50\%$), and to make the recall of the membership inference model as low as possible.

\textbf{Metrics for Evaluating the Target Model:}
To further demonstrate that the proposed method does not affect the performance of the target model, we also compare the accuracy of the target model with and without defense on the training set and test set.
The \textit{training accuracy} is the proportion of training data that are correctly classified by the target model \cite{shalev2014understanding}.
The \textit{test accuracy} is the proportion of test data that are correctly classified by the target model \cite{shalev2014understanding}.
The goal is to ensure that the proposed defense method will not affect the training accuracy and test accuracy of the target model.

\textbf{Metrics for Evaluating the Similarity Between the Original Prediction and the Adversarial Prediction:}
We use the size of the added perturbations to represent the similarity between the original prediction and the adversarial prediction. The size of the added perturbations is calculated as follows \cite{carlini2017adversarial}: ${\left\| \delta  \right\|_1} = \sum\nolimits_{i = 0}^n {\left\| {{\delta _i}} \right\|}$.
The smaller the value of $\left\| \delta  \right\|_1$ is, the smaller the added adversarial perturbation on each confidence score of a prediction.
In other words, the difference between the original prediction and the adversarial prediction is more imperceptible.

\section{The Proposed AEPPT}\label{sec:The_proposed_AEPPT}

\subsection{Threat Model}\label{sec:threat_model}

If a defense method can successfully resist membership inference attacks in the worst case, then the method can provide effective and strong protection for the target model. As defenders, they need to resist the most capable attackers.
Therefore, in this work, we assume that the adversary has strong capabilities in two aspects: the adversary's knowledge, and the adversary's data.

The adversary's knowledge about the target model is as follows: the adversary knows the structure and the training algorithm of the target model \cite{salem2018ml}.
In a common scenario for current MLaaS, the adversary can only query the target model and obtain the prediction \cite{nasr2018machine}.

The training data used by the target model is denoted as $D_{tar}$, and the adversary's data is denoted as $D_{adv}$. $\mathfrak{D}$ represents the distribution of the data. The relationship between the adversary's data and the training data of the target model is as follows:

\begin{itemize}
  \item { ${D_{tar}} \cap {D_{adv}} = {D_{adv}}  $: The adversary knows part of the training data of the target model.}
  \item {${D_{tar}} \sim \mathfrak{D}$, ${D_{adv}} \sim \mathfrak{D}$: The adversary's data is from the same distribution as the training data of the target model \cite{yeom2018privacy}.}
\end{itemize}

In this paper, we consider a more complex inference attack scenario than the black-box settings, where the adversary's data $D_{adv}$ is entirely from the training data $D_{tar}$ of the target model, \textit{i.e.}, ${D_{tar}} \cap {D_{adv}} = {D_{adv}}$. In this way, the attacker can train the membership inference model better to implement more powerful membership inference attacks. The experiments in Section \ref{sec:exp_res} show that, the proposed AEPPT method is still effective under such complex attack scenario.

\subsection{Overall Flow of the Proposed Method}

To resist against membership inference attacks, AEPPT needs to ensure that the generated adversarial prediction can mislead the membership inference model successfully. However, the membership inference model is generally a black-box model for the defender in practice, \textit{i.e.}, the defender does not have the knowledge about the adversary's membership inference model.
To solve this problem, before generating adversarial prediction, the AEPPT first constructs a substitute membership inference model to mimic the behavior of the attacker's membership inference model.
Then, the adversarial prediction is generated based on the substitute membership inference model, which can mislead the adversary's membership inference model and resist membership inference attacks.

\begin{figure}[htbp]
    \includegraphics[width=3.5in]{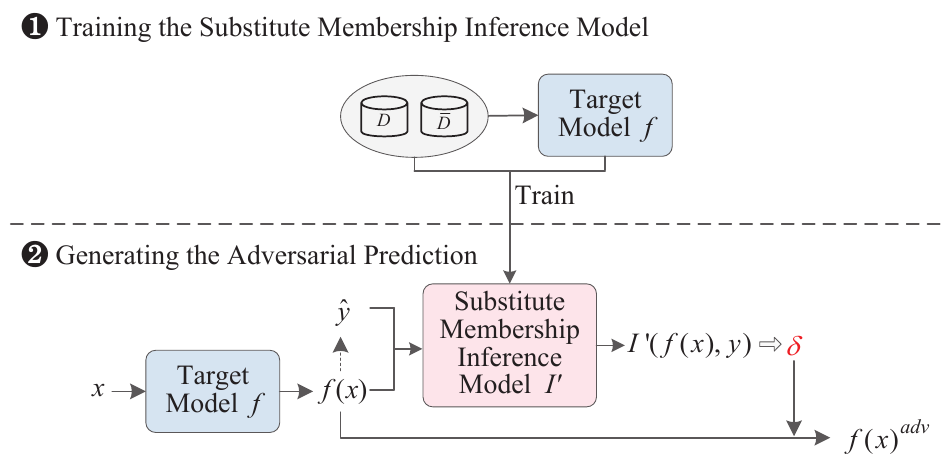}\\
  \caption{Overall flow of the proposed AEPPT.}
  \label{fig:ODMIA}
\end{figure}

The overall flow of the proposed AEPPT is illustrated in Fig. \ref{fig:ODMIA}.
In the first step, the substitute membership inference model is trained.
AEPPT first obtains the predictions of member data and non-member data from the target model. Then, member data, non-member data, and corresponding predictions constitute the new training data which is used to train the substitute membership inference model.
In the second step, the adversarial prediction is generated. When the target model $f$ receives a given input $x$, the target model first calculates the prediction result $f (x)$. Then, the prediction $f(x)$ is converted to a label $\hat y$ using one-hot encoding \cite{harris2010digital}, \textit{i.e.}, the maximum component of $f(x)$ is set to be 1, and the other components in $f(x)$ are set to be 0. After that, the prediction result $f(x)$ of the target model and the converted label $\hat y$ are input to the substitute membership inference model $I'$. The substitute membership inference model calculates the probability of that the data $(x,y)$ is in the training set of the target model, which is denoted as $I'(f(x),\hat y)$.
According to the output $I'(f(x),\hat y)$ of the substitute membership inference model, AEPPT crafts perturbations $\delta$ and adds them to the output $f(x)$ of the target model to generate the adversarial prediction $f(x)^{adv}$. Finally, the adversarial prediction $f(x)^{adv}$ is used as the new output of the target model.

The proposed method of training the substitute membership inference model is described in Section \ref{sec:SMIMG}, and the proposed method of generating the adversarial prediction against membership inference attacks is presented in Section \ref{sec:APG}.

\subsection{Training the Substitute Membership Inference Model}\label{sec:SMIMG}

\begin{figure}[htbp]
  \centering
  \includegraphics[width=3in]{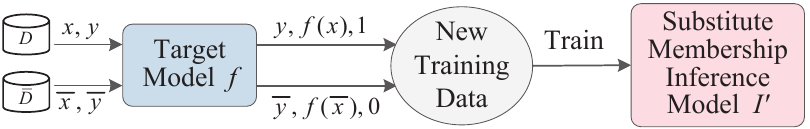}
  \caption{Training process of the substitute membership inference model.}
  \label{fig:TPSMIM}
\end{figure}

Fig. \ref{fig:TPSMIM} shows the training process of the substitute membership inference model. First, the training data of the substitute membership inference model is constructed. The training data of the substitute membership inference model consists of two types of data: member data $D$ and non-member data $\bar D$.
Member data $D$ is the data that has been used to train the target model, while non-member data $\bar D$ is the data that has not been used to train the target model \cite{shokri2017membership}.
In order to avoid a low accuracy of the substitute membership inference model due to the large difference between the number of member data and the number of non-member data, the number of member data should be equal to the number of non-member data.
The training data of the substitute membership inference model is constructed as follows.
For a record $(x,y) \in D$, the target model outputs the prediction result $f(x)$.
Then, the label $y$, the prediction result $f(x)$, and the integer $1$ are combined to form a training data $(y,f(x),1)$ of the substitute membership inference model.
Similarly, for a record $(\bar x, \bar y) \in \bar D$, the target model outputs the prediction result $f(\bar x)$.
The label $\bar y$, the prediction result $f (\bar x)$, and the integer $0$ are combined to form a training data $(\bar y,f(\bar x),0)$ of the substitute membership inference model.
The integer 1 indicates that the data $x$ is a member data, while integer 0 means that $x$ is a non-member data.

\begin{figure}[htbp]
  \centering
  \includegraphics[width=2.5in]{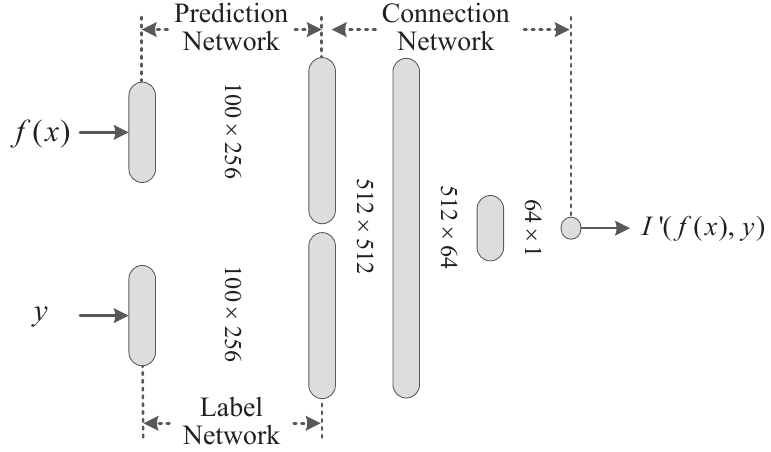}
  \caption{Structure of the substitute membership inference model \cite{nasr2018machine}.}
  \label{fig:AMIM}
\end{figure}

In this work, the structure of the substitute membership inference model is similar to the structure of the membership inference model used in \cite{nasr2018machine}, which is composed of three separate fully connected sub-networks \cite{nasr2018machine}: prediction network, label network, and connection network. The input of the prediction network is the prediction vector of the target model, and the input of the label network is the true label $y$ \cite{nasr2018machine}. The input of the connection network is the output of prediction network and label network, and the output of the connection network is the probability of that the data $x$ is in the training set of the target model \cite{nasr2018machine}.
Compared with the membership inference model used in \cite{nasr2018machine}, the layer sizes of the three sub-networks in the substitute membership inference model are different from that in \cite{nasr2018machine}. The structure of the substitute membership inference model is illustrated in Fig. \ref{fig:AMIM}. The layer sizes of the prediction network, label network, and connection network in the substitute membership inference model are [100, 256], [100, 256], and [512, 64, 1], respectively.

Since the substitute membership inference model mimics the behavior of the adversary's membership inference model, intuitively, the defensive performance of the adversarial prediction may be affected by the substitute membership inference model. To demonstrate that the generated adversarial prediction is effective when facing different membership inference models, we use membership inference models that are different from the substitute membership inference model to evaluate the proposed method in Section \ref{sec:exp_diff_MIM}.

\subsection{Generating the Adversarial Prediction against Membership Inference Attacks}\label{sec:APG}

We propose an adversarial prediction generation algorithm, as shown in Algorithm \ref{alg:APGA}. Specifically, based on the adversarial examples generation method, such as \cite{goodfellow2014explaining,kurakin2016adversarial,carlini2017towards}, the proposed adversarial prediction generation algorithm modifies the prediction of target model to resist membership inference attacks. First, the algorithm converts the target model's original prediction $f(x)$ to $\hat y$, and computes the output of the substitute membership inference model. Then, the adversarial prediction $f{(x)^{adv}}$ is initialized. After that, the perturbations are added to $f{(x)^{adv}}$ by iterative optimization to generate the final adversarial prediction. The process of optimizing the added perturbations is described as follows.

\renewcommand{\algorithmicrequire}{\textbf{Input:}}
\renewcommand{\algorithmicensure}{\textbf{Output:}}
\begin{algorithm}
\caption{Adversarial Prediction Generation Algorithm}
\label{alg:APGA}
\begin{algorithmic}[1]
\REQUIRE target model's prediction $f(x)$, substitute membership inference model $I'$, the step size $\epsilon $, the number of iterations $T$
\ENSURE adversarial prediction $f(x)^{adv}$
\STATE Convert $f(x)$ to $\hat y$
\STATE Compute the output of the substitute membership inference model: $I'(f{(x)},\hat y)$
\STATE Initialize $f{(x)^{adv}}$: $f{(x)_{0}^{adv}} \leftarrow f(x)$
\FOR {$t=0$ to $T-1$}
    \STATE Compute the output of the substitute membership inference model: $I'(f{(x)_{t}^{adv}},\hat y)$
    \IF {$I'(f(x),\hat y) \ge {\rm{0}}{\rm{.5}}$}
        \STATE ${\delta _t} \leftarrow sign[{\nabla _{f(x)}}J(I'(f(x)_t^{adv},\hat y),1)]$
    \ELSE
        \STATE ${\delta _t} \leftarrow sign[{\nabla _{f(x)}}J(I'(f(x)_t^{adv},\hat y),0)]$
    \ENDIF
    \STATE Determine the random step size $\epsilon '$: \\~~~~~~$\epsilon ' \leftarrow random(0,1) \cdot \epsilon $
    \STATE Update $f(x)_{t+1}^{adv}$: $f(x)_{t+1}^{adv} \leftarrow f(x)_t^{adv} + \epsilon ' \cdot {\delta _t}$
    \STATE $f(x)_{t+1}^{adv} \leftarrow Clip\{ f(x)_{t+1}^{adv},0,1\} $
\ENDFOR
\STATE $f(x)^{adv} \leftarrow f(x)_{T}^{adv}$
\RETURN  $f(x)^{adv}$
\end{algorithmic}
\end{algorithm}

Different from the adversarial examples generation methods, \textit{e.g.}, \cite{goodfellow2014explaining,kurakin2016adversarial,carlini2017towards}, which add perturbations to the input data of the target model, the proposed method generates and adds perturbations to the output of the target model.
Specifically, the prediction $f(x)$ of target model contains $N$ components, each of which represents the confidence score of a different class label ($N$ classes in total).
The proposed defense method adds the perturbation on each component of the prediction $f(x)$.
In other words, the generated perturbations will be distributed on the $N$ confidence scores of the original prediction $f(x)$ to construct the adversarial prediction $f{(x)^{adv}}$.
Based on the basic iterative method \cite{kurakin2016adversarial}, the adversarial prediction can be calculated as follows \cite{kurakin2016adversarial}:
\begin{equation}
  f(x)_0^{adv} = f(x),~~~f(x)_{t + 1}^{adv} = Clip\{ f(x)_t^{adv} + \epsilon  \cdot {\delta _t}\}
\end{equation}
where $t$ represents the $t$-th round of iteration, and $\epsilon$ is the size of added perturbations in each iteration.
To prevent the added perturbations from being too large, a clipping function $Clip\{\}$ \cite{kurakin2016adversarial} is used to constrain the value of each element in the generated adversarial prediction to between $0$ and $1$. The added perturbations are calculated as follows \cite{kurakin2016adversarial}:
\begin{equation}
  {\delta _t} = sign[{\nabla _{f(x)}}J(I'(f(x)_t^{adv},\hat y),Y)]
\end{equation}
where $J(\cdot)$ is a loss function, and ${\nabla _{f(x)}}J(\cdot)$ is the gradient of the loss function. $Y$ indicates whether $x$ is in the training set of the target model or not. If $Y$ is 1, it means that $x$ is in the training set of the target model. If $Y$ is 0, it means that $x$ is not in the training set of the target model.

For the output $I'(f(x),\hat y)$ of the substitute membership inference model, if $I'(f(x),\hat y)$ is greater than or equal to 0.5, it represents that the data $(x,y)$ is in the training set of the target model. Otherwise, the data $(x,y)$ is not in the training set of the target model.
To ensure that the generated adversarial prediction can fool the membership inference model successfully, the added perturbations can be optimized according to the prediction result of the substitute membership inference model.
Specifically, if the data $(x,y)$ is in the training set of the target model originally, the goal of the added perturbations is to make the attacker infer that the data $(x,y)$ is not in the training set of the target model. On the contrary, if the data $(x,y)$ is not in the training set of the target model originally, the goal of the added perturbations is to make the attacker infer that the data $(x,y)$ is in the training set of the target model. Therefore, in the proposed method, the added perturbations are generated as follows:
\begin{equation}
\begin{split}
\delta_{t} &=
\begin{cases}
sign[{\nabla _{f(x)}}J(I'(f(x)_t^{adv},\hat y),1)],&I'(f(x),\hat y) \ge 0.5 \\
sign[{\nabla _{f(x)}}J(I'(f(x)_t^{adv},\hat y),0)],&I'(f(x),\hat y) < 0.5
\end{cases}
\end{split}
\label{flip_label}
\end{equation}

In the adversarial prediction generation process, using the same step size $\epsilon$ and the same number of iterations may produce the same components in different adversarial predictions. As a result, an attacker may suspect that perturbations are deliberately added to the prediction of the target model.
In this paper, to prevent the adversary from finding the added perturbations in a generated adversarial prediction, we add the random size perturbation into the prediction at each iteration process, so that the perturbations on each adversarial prediction $f(x)^{adv}$ are different.
Specifically, in each iteration, a random number between $0$ and $1$ is generated, and then the random number is multiplied by the input step size $\epsilon$ to obtain a random step size $\epsilon'$. The generated perturbations $\delta$ multiplied by the random step size $\epsilon'$ are added to $f(x)^{adv}$. Thus, the size of the added perturbation in each iteration is variable.
After multiple iterations, even if the two original predictions are the same, there will be a slight difference between the final generated adversarial predictions.
In this way, even the attacker queries the same input data for multiple times, the adversarial perturbations added on the prediction at each query will be different.
In other words, except for the predicted label, the attacker cannot gain any additional information from those generated adversarial predictions.

\section{Experimental Evaluation}\label{sec:Experiments}
First, datasets and models used in the experiment are described in Section \ref{sec:dataset} and Section \ref{sec:MIM_Intro}, respectively. Then, the performance of the proposed method is evaluated on the adversary's membership inference model and the target model in Section \ref{sec:exp_res}.
The factors that may affect the defensive performance are analyzed in Section \ref{sec:exp_diff_para}$\sim$Section \ref{sec:exp_diff_MIM}, respectively.
The defense performance under three adaptive membership inference attacks are evaluated in Section \ref{sec:adaptive}.
Section \ref{sec:discussion} compares the proposed method with the state-of-the-art defense works.

\subsection{Dataset}\label{sec:dataset}

\textbf{CIFAR100} \cite{krizhevsky2009learning}: This dataset contains 60,000 images (50,000 training images and 10,000 test images), and the size of each image is $32\times32$ \cite{krizhevsky2009learning}. The images in the dataset are classified into 100 classes, such as fish, chair, cloud, bicycle, etc. Each class contains 500 training images and 100 test images. In the experiment, we divide this dataset according to the adversary's knowledge, and then use these data to train the target model and membership inference model, respectively.

\textbf{Purchase} \cite{kaggle_challenge}: This dataset is constructed based on the data from the ``acquire valued shoppers'' challenge on Kaggle \cite{kaggle}. The goal of this challenge is to find customers who are most likely to repeat purchase according to the purchase history. Because the raw data provided by this challenge contains a lot of redundant information, such as the number of the product purchase, the date of purchase, and the amount of the purchase, we use the same method as in \cite{shokri2017membership} to simplify the dataset. The simplifying process is as follows. Each record with 600 binary features represents a customer, and each feature represents a product \cite{shokri2017membership}. If the customer has purchased the product, the value of the corresponding feature is set to be 1. Otherwise, it is set to be 0. Then, we cluster these records into 100 classes, and set the corresponding label for each record \cite{shokri2017membership}. Finally, we use the simplified data to train the target model and membership inference model, respectively.

\subsection{Target Model and Membership Inference Model}\label{sec:MIM_Intro}

\textbf{Target Model:} For CIFAR100 dataset, we use the AlexNet architecture proposed in \cite{krizhevsky2012imagenet} to train the target model. We use the cross-entropy loss function to train the target model based on Pytorch\footnote{\url{https://pytorch.org}}, and set the learning rate and the maximum epochs of training to be 0.0001 and 100, respectively.
For the Purchase dataset, we use the same fully connected neural network as in \cite{nasr2018machine} to train the target model. The layer size and activation function of the fully connected neural network are the same as those in \cite{nasr2018machine}, which are [1024, 512, 256, 100] and $Tanh$. We train the target model with cross-entropy loss function, and set the learning rate and the maximum epochs of training to be 0.0001 and 100, respectively.
We divide the data in the dataset into training data $D_{tar}$ and non-training data $\bar{D}_{tar}$, and use training data to train the target model. The number of training data and non-training data on the CIFAR100 and Purchase datasets are shown in Table \ref{tab:member_non_data}.

\begin{table}[htbp]
  \centering
  \footnotesize
  \caption{The Number of Training Data $D_{tar}$ and Non-training Data $\bar{D}_{tar}$  on the CIFAR100 and Purchase Datasets.}
    \begin{tabular}{|c|c|c|}
    \hline
    Dataset &  $D_{tar}$  & $\bar{D}_{tar}$  \\
    \hline
    CIFAR100 & 50,000 & 10,000 \\
    \hline
    Purchase & 20,000 & 20,000 \\
    \hline
    \end{tabular}%
  \label{tab:member_non_data}%
\end{table}%

\textbf{Substitute Membership Inference Model:} The structure of the substitute membership inference model is described in Section \ref{sec:SMIMG}. The activation function of the substitute membership inference model is $ReLU$ (Rectified Linear Unit) \cite{NairH10}. We use the cross-entropy loss function \cite{li1998iterative} to train the substitute membership inference model, and set the learning rate and the maximum epochs of training to be 0.001 and 100, respectively.
The training data of the substitute membership inference model consists of member data (denoted as $D_{sub}$) and non-member data (denoted as $\bar{D}_{sub}$). We use training data $D_{tar}$ and non-training data $\bar{D}_{tar}$ as member data $D_{sub}$ and non-member data $\bar{D}_{sub}$, respectively. The number of member data $D_{sub}$ is equal to the number of training data (\textit{i.e.}, {\small $\left| D_{sub} \right| = \left| D_{tar} \right|$}), and the number of non-member data $\bar{D}_{sub}$ is equal to the number of non-training data (\textit{i.e.}, {\small $\left| \bar{D}_{sub} \right| = \left| \bar{D}_{tar} \right|$}).

\textbf{Adversary's Membership Inference Model:}
In the experiment, the structure of the adversary's membership inference model is the same as that in \cite{nasr2018machine}. For simplicity, we denote this membership inference model as MIM0. The layer sizes of three sub-networks and activation function of MIM0 are the same as those in \cite{nasr2018machine}, which are [100, 1024, 512, 64], [100, 512, 64], [128, 256, 64, 1] and $ReLU$ \cite{NairH10}, respectively.
We use the cross-entropy loss function \cite{li1998iterative} to train the adversary's membership inference model, and set the learning rate and the maximum epochs of training to be 0.001 and 100, respectively.
The training data of the adversary's membership inference model consists of member data (denoted as $D_{adv}$) and non-member data (denoted as $\bar{D}_{adv}$).
As discussed in Section \ref{sec:threat_model}, this paper assumes that the attacker has strong knowledge about the training data of the target model.
Specifically, we randomly select data from training data $D_{tar}$ and non-training data $\bar{D}_{tar}$ as member data $D_{adv}$ and non-member data $\bar{D}_{adv}$, respectively.
The number of member data $D_{adv}$ is half of the training data (\textit{i.e.}, {\small $\left| D_{adv} \right| = \frac{1}{2} \left| D_{tar} \right|$}), and the number of non-member data $\bar{D}_{adv}$ is half of the non-training data (\textit{i.e.}, {\small $\left| \bar{D}_{adv} \right| =  \frac{1}{2} \left| \bar{D}_{tar} \right|$}).

\subsection{Experimental Results} \label{sec:exp_res}

\begin{figure*}[htbp]
  \centering
  \subfigure[CIFAR100 Dataset]{
    \label{}
    \includegraphics[width=7in]{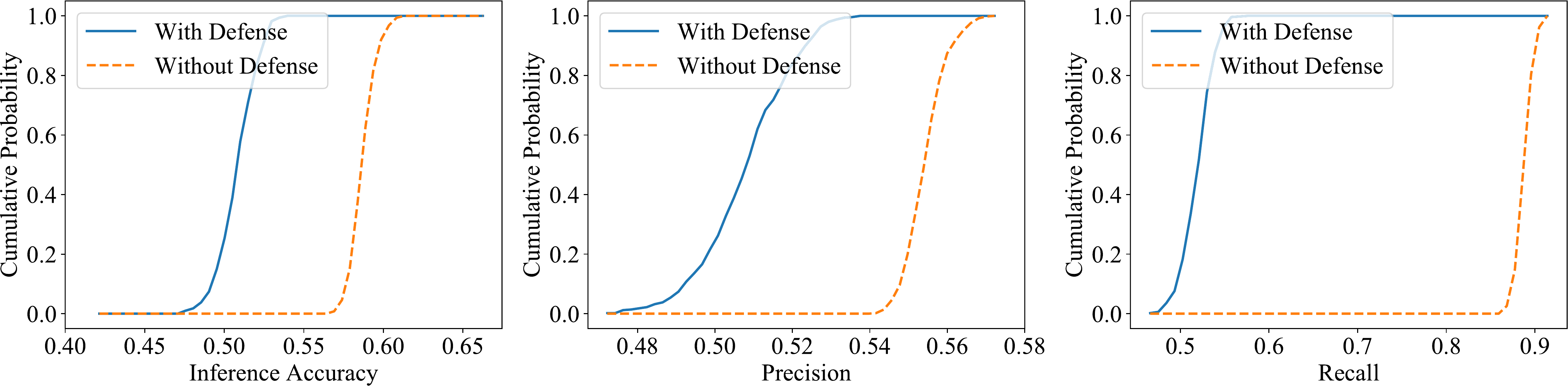}}
      \subfigure[Purchase Dataset]{
    \label{}
    \includegraphics[width=7in]{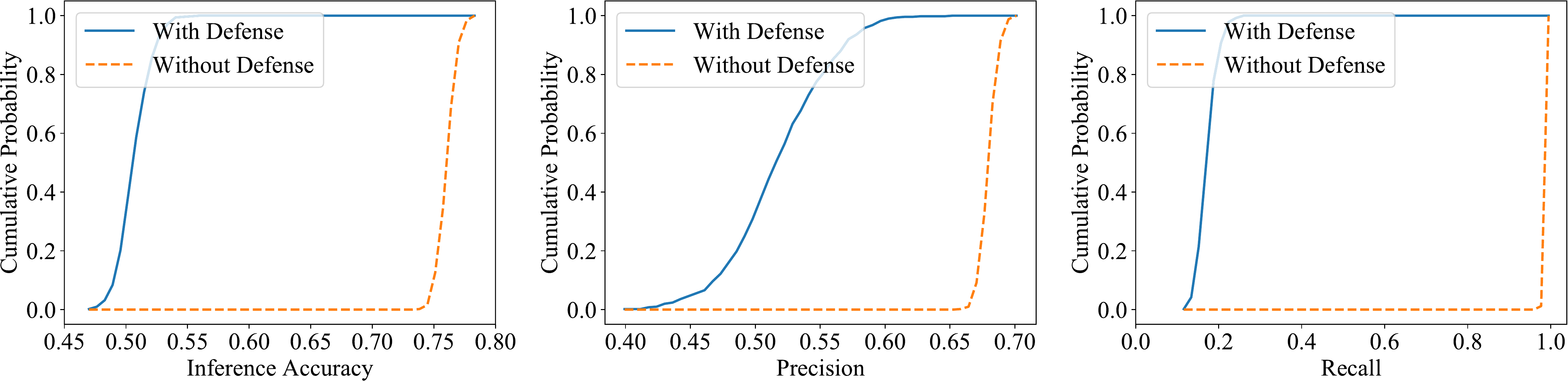}}
  \caption{Empirical CDF of inference accuracy, precision, and recall of the membership inference model on the CIFAR100 and Purchase datasets. }
  \label{fig:cdf_attack}
\end{figure*}

\textbf{Evaluation on the Membership Inference Model:}
In the experiment, for each output class of the target model, we evaluate the inference accuracy, precision, and recall of the membership inference model, and obtain the empirical cumulative distribution function (CDF) \cite{gentle2009computational} of the inference accuracy, precision, and recall. For a given value $w$, the empirical CDF represents the proportion of all results $(w_{1},w_{2},...,w_{m})$ that are less than or equal to the value $w$, which can be calculated as follows \cite{gentle2009computational}:
\begin{equation}
  P_{m}(w)=\frac{\#\{w_{i},~s.t.~w_{i}<w\}}{m}
\end{equation}
where $w_1,w_2,...,w_m$ represent the results of $m$ times.
Fig. \ref{fig:cdf_attack} shows the empirical CDF of inference accuracy, precision, and recall of the membership inference model on the CIFAR100 and Purchase datasets, respectively.
The results show that the proposed defense method can make the inference accuracy of membership inference model reduce to that of a random guess, which means the proposed method can make the membership inference attack completely fail. For example, the proposed method reduces the inference accuracy of the membership inference model from $76\%$ to $51\%$ on the Purchase dataset, and reduces from $59\%$ to $52\%$ on the CIFAR100 dataset. Under the protection of the proposed method, the precision and recall of the adversary's membership inference model are also greatly reduced. For example, on the Purchase dataset, the recall of the membership inference model against the target model without defense is higher than $96\%$, while the recall of the membership inference model against the target model with defense is lower than $25\%$.
Therefore, the proposed defense method can effectively resist membership inference attacks, and protect the training data of the target model.

\begin{figure}[htbp]
  \centering
  \includegraphics[width=2.5in]{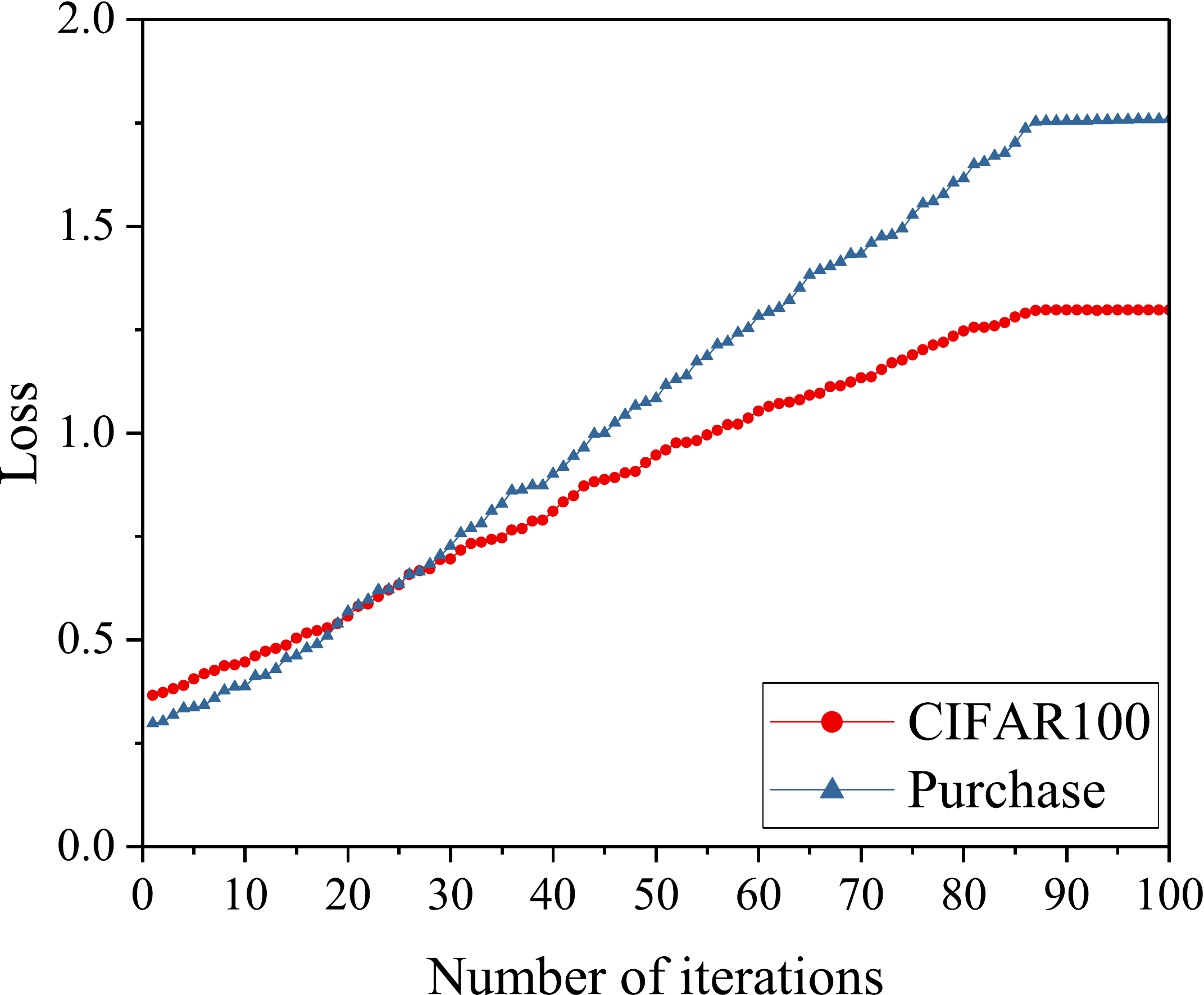}
  \caption{The value of loss function of the adversary's membership inference model during the generation of adversarial prediction.}
  \label{fig:loss}
\end{figure}

Fig. \ref{fig:loss} shows the value of loss function of the adversary's membership inference model during the adversarial prediction generation. It is shown that the generated adversarial prediction can affect the loss function of the attacker's membership inference model. The more iterations, the larger the loss function.
The larger the loss function, the lower the inference accuracy of the membership inference model.
After multiple rounds of optimizations, the loss function of the adversary's membership inference model has stabilized. Finally, the generated adversarial prediction can prevent the attacker from correctly inferring whether a specific data is in the training set of the target model.

\textbf{Evaluation on the Target Model:}
We train the target model on the CIFAR100 dataset and Purchase dataset, respectively. On the CIFAR100 dataset, the target model has stabilized after 75 rounds of iterative training. On the Purchase dataset, the target model has stabilized after 50 rounds of iterative training.
Table \ref{tab:tarmodel_acc} shows the training accuracy and test accuracy of the target model with and without defense on the CIFAR100 and Purchase datasets.
It is shown that, the proposed method will not affect the normal performance of target model, where the fluctuation of the training accuracy and test accuracy is as low as 0.1\%.
Such an accuracy fluctuation is common when training the same DNN model for multiple times, which is negligible.
As discussed in Section \ref{sec:APG}, the proposed method does not modify or retrain the target model, but only adds small perturbations to the original prediction of the target model to generate an adversarial prediction.
In this way, the generated prediction does not affect the accuracy of the target model, but can effectively mislead membership inference models.

\begin{table}[htbp]
  \centering
  \footnotesize
  \caption{Training Accuracy and Test Accuracy of the Target Model with and without Defense on CIFAR100 and Purchase Datasets.}
    \begin{tabular}{|c|c|c|c|c|}
    \hline
    \multicolumn{1}{|c|}{\multirow{2}[9]{*}{Dataset}} & \multicolumn{2}{c|}{CIFAR100} & \multicolumn{2}{c|}{Purchase} \\

\cline{2-5}    \multicolumn{1}{|c|}{} & \multicolumn{1}{p{4.19em}|}{\tabincell{c}{Training \\Accuracy}} & \multicolumn{1}{c|}{\tabincell{c}{Test \\Accuracy}} & \multicolumn{1}{c|}{\tabincell{c}{Training \\Accuracy}} & \multicolumn{1}{c|}{\tabincell{c}{Test \\Accuracy}} \\
    \hline
    Without Defense & 96.1\% & 50.2\% & 99.3\% & 79.7\% \\
    \hline
    \textbf{With Defense} & \textbf{96.0\%} & \textbf{50.2\%} & \textbf{99.2\%} & \textbf{79.4\%} \\
    \hline
    \end{tabular}%
  \label{tab:tarmodel_acc}%
\end{table}%

\subsection{Defensive Performance with Different Perturbation Step Size}\label{sec:exp_diff_para}

In the proposed method, the size of the added perturbations in the prediction of the target model has an impact on the defensive performance. To this end, we evaluate the inference accuracy, precision, and recall of the membership inference model under different perturbation step sizes.

Table \ref{tab:cifar_para} and Table \ref{tab:purcahse_para} show the size of the added perturbations and the inference accuracy of the membership inference model under different perturbation step sizes on the CIFAR100 and Purchase datasets, respectively.
As shown in Fig. \ref{fig:loss}, the loss function of the adversary's membership inference model is stable when the adversarial perturbations has been optimized for about 100 rounds.
This indicates that the proposed AEPPT method has already reduced the accuracy of membership inference attacks to a stable value (around 50\%).
Therefore, in our experiments, the number of iterations to generate the adversarial perturbations is set to be 100.

\begin{table}[htbp]
  \centering
  \footnotesize
  \caption{Size of Added Perturbations and the Inference Accuracy of the Membership Inference Model under Different Perturbation Step Sizes on the CIFAR100 Dataset.}
    \begin{tabular}{|c|c|c|}
    \hline
    \tabincell{c}{Perturbation \\Step Size $\epsilon$} & \tabincell{c}{Size of Added\\Perturbations} & \tabincell{c}{Inference \\Accuracy } \\
    \hline
    0 & 0& 58.7\% \\
    \hline
    $7 \times 10^{-4}$ &0.1842& 52.2\% \\
    \hline
    $8 \times 10^{-4}$ &0.2208& 51.4\%  \\
    \hline
    $9 \times 10^{-4}$ &0.2637& 50.3\%  \\
    \hline
    $1 \times 10^{-3}$ &0.2961& 49.1\%  \\
    \hline
    \end{tabular} \\
    \footnotesize{$^*$ The size of added perturbations is the sum of perturbations that added on the confidence score of all classes (100 in total), and the average perturbation on each confidence score is as low as 0.001842.}
  \label{tab:cifar_para}%
\end{table}%

\begin{table}[htbp]
  \centering
  \footnotesize
  \caption{Size of Added Perturbations and the Inference Accuracy of the Membership Inference Model under Different Perturbation Step Sizes on the Purchase Dataset.}
    \begin{tabular}{|c|c|c|}
    \hline
    \tabincell{c}{Perturbation \\Step Size $\epsilon$} &\tabincell{c}{Size of Added\\ Perturbations} & \tabincell{c}{Inference \\Accuracy} \\
    \hline
    0    & 0 & 76.0\%  \\
    \hline
    $1\times 10^{-5}$&0.0485 & 74.2\% \\
    \hline
    $1.5\times 10^{-5}$&0.0736 & 61.7\%  \\
    \hline
    $2\times 10^{-5}$&0.0868 & 50.6\% \\
    \hline
    $2.5\times 10^{-5}$&0.0882 & 47.4\%  \\
    \hline
    \end{tabular}\\
    \footnotesize{$^*$ The size of added perturbations is the sum of perturbations that added on the confidence score of all classes (100 in total), and the average perturbation on each confidence score is as low as 0.000485.}
  \label{tab:purcahse_para}%
\end{table}%

The $L_{1}$ distance is used to calculate the size of the added perturbations, \textit{i.e.},\scriptsize{ $\left\| {f{{(x)}^{adv}} - f(x)} \right\|_{1}$} \normalsize\cite{carlini2017adversarial}. It is shown in Table \ref{tab:cifar_para} and Table \ref{tab:purcahse_para} that, the proposed method only needs to add small perturbations to the original prediction to make the inference accuracy of the membership inference model reduce to about 50\%. As discussed in Section \ref{sec:APG}, the generated adversarial perturbations are distributed on each component (\textit{i.e.}, each confidence score) of the original prediction.
For instance, when the perturbation step size is set to be $2\times 10^{-5}$, the size of the perturbations on the adversarial prediction is 0.0868 on the Purchase dataset.
The average modifications on each component (100 components in total) is only 0.000868.
The larger the added perturbations is, the lower the inference accuracy of the membership inference model. The reason is that a large perturbation step size makes the added perturbations in the adversarial prediction large, and the generated adversarial prediction is more likely to mislead the membership inference model. The goal of the proposed method is to make the inference accuracy and precision of the membership inference model reduce to around $50\%$. Therefore, for CIFAR100 and Purchase datasets, the perturbation step size $\epsilon$ is set to be $9 \times 10^{-4}$ and $2\times 10^{-5}$, respectively, which ensures that the accuracy of membership inference model is close to a random guess.

\begin{figure}[htbp]
  \centering
  \includegraphics[width=3.50in]{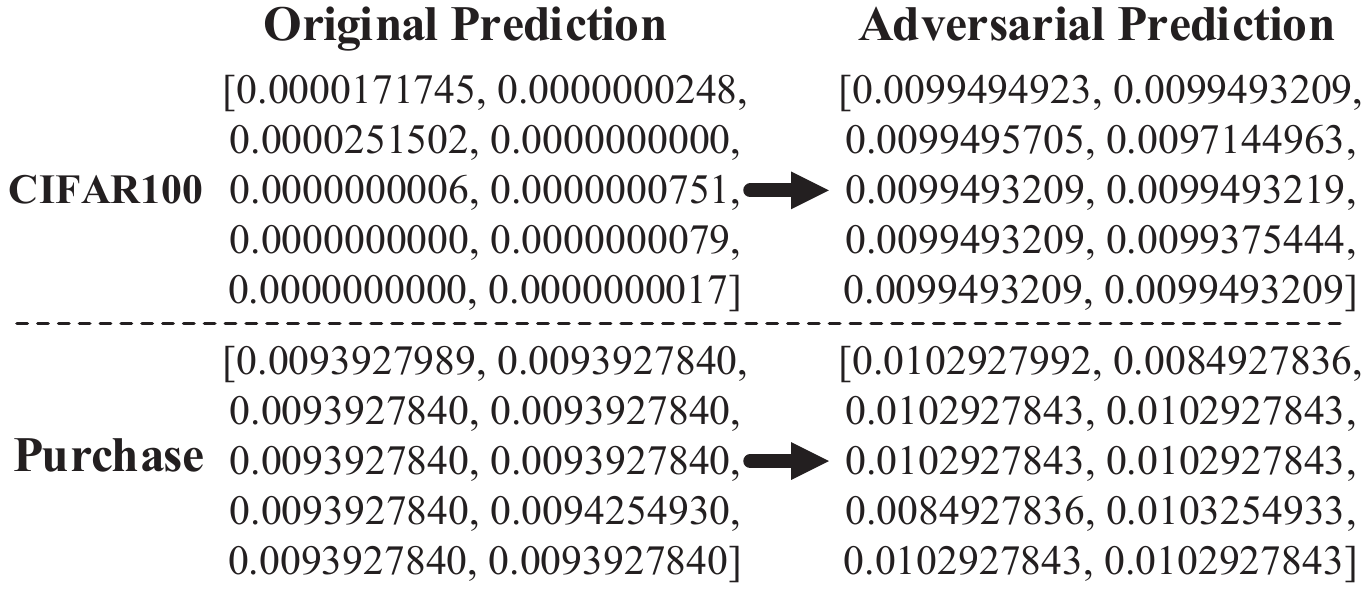}
  \caption{An example of the original prediction and the generated adversarial prediction on CIFAR100 and Purchase datasets, respectively.}
  \label{fig:comprason}
\end{figure}

Fig. \ref{fig:comprason} presents an example of the original prediction and the generated adversarial prediction on CIFAR100 and Purchase datasets, respectively.
For simplicity, we only shows the confidence scores of the first 10 classes on each prediction.
It is shown that the added perturbation on the confidence score of each class is small, and the size of perturbation on each component is in the order of $10^{-3}$.
Therefore, the proposed defense will not affect the utility of the final prediction of the target model.

\subsection{Effect of the Number of the Target Model's Output Classes}\label{sec:exp_diff_class}

\begin{table*}[htbp]
  \centering
  \footnotesize
  \caption{Inference Accuracy, Precision, and Recall of the Adversary's Membership Inference Model against Target Models with Different Numbers of Output Classes on Purchase Dataset.}
    \begin{tabular}{|c|c|c|c|c|c|c|c|c|}
    \hline
          & \multicolumn{2}{c|}{Target Model} & \multicolumn{3}{c|}{Without Defense} & \multicolumn{3}{c|}{With Defense} \\
    \hline
    \tabincell{c}{Number of \\Classes} & \tabincell{c}{Training \\Accuracy} & \tabincell{c}{Test \\Accuracy} & \tabincell{c}{Inference \\Accuracy} & Precision & Recall & \tabincell{c}{Inference \\Accuracy} & Precision & Recall \\
    \hline
    2     & 99.3\% & 98.9\% & 49.5\% & 49.5\% & 47.9\% & 50.8\% & 50.8\% & 50.5\% \\
    \hline
    10    & 99.6\% & 95.2\% & 54.5\% & 52.4\% & 97.2\% & 49.6\% & 49.4\% & 33.0\% \\
    \hline
    20    & 99.8\% & 93.0\% & 61.1\% & 57.3\% & 88.0\% & 52.7\% & 53.4\% & 42.3\% \\
    \hline
    50    & 99.4\% & 86.6\% & 69.2\% & 62.0\% & 99.4\% & 48.0\%  & 47.2\% & 33.0\% \\
    \hline
    100   & 99.3\% & 79.7\% & 76.0\% & 68.0\% & 98.5\% & 50.6\% &  51.7\% & 17.1\% \\
    \hline
    \end{tabular}%
  \label{tab:diff_class}%
\end{table*}%

In \cite{shokri2017membership}, the authors indicate that the more the number of the target model's output classes, the more information the adversary can obtain from the target model, and the higher the success rate of the membership inference attack. In order to evaluate the effect of the number of the target model's output classes on the defensive performance of the proposed method, we cluster the data in Purchase dataset into 2, 10, 20, 50, 100 classes \cite{shokri2017membership}, respectively.
Then, we train target models with different number of output classes respectively \cite{shokri2017membership}, and evaluate the inference accuracy, precision, and recall of the membership inference model.

Table \ref{tab:diff_class} shows the inference accuracy, precision, and recall of the membership inference model against target models with different number of output classes on Purchase dataset. For target models without defense, the more the number of the target model's output classes, the higher the inference accuracy of the membership inference model. For instance, when the number of the target model's output classes is only two, the inference accuracy and precision of the membership inference model are around $50\%$. When the number of the target model's output classes is $100$, the inference accuracy and precision of the membership inference model are $76\%$ and $68\%$, respectively. For target models with defense, the inference accuracy and precision of the membership inference model against target models with different numbers of output classes are all close to that of a random guess. In other words, the proposed method can effectively resist membership inference attacks regardless of the number of the target model's output classes.

\subsection{Effect of the Number of the Adversary's Data}\label{sec:exp_diff_data}
\begin{figure*}[htbp]
  \centering
  \includegraphics[width=6.9in]{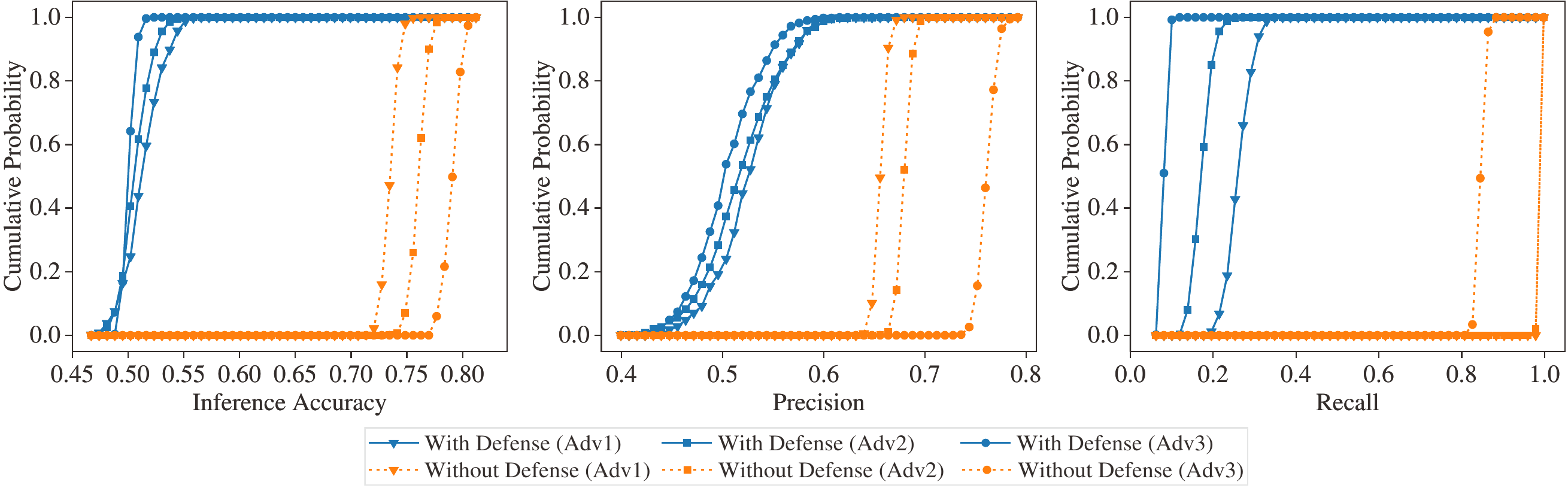}
  \caption{Empirical CDF of inference accuracy, precision, and recall of three different adversaries' membership inference models on the Purchase dataset.}
  \label{fig:adv123}
\end{figure*}

The adversary's data also affects the performance of membership inference model. In the above experiments, the number of adversary's data is set to be half of the number of the training data of the target model. To evaluate the effect of the number of adversary's data on the performance of the proposed method, three adversaries with different numbers of data are evaluated: adversary 1 (Adv1), adversary 2 (Adv2), and adversary 3 (Adv3).
The numbers of the three adversaries' data are set as follows:
\begin{itemize}
\setlength{\baselineskip}{20pt}
  \item {
  \textit{Adv1:} \small $\left| {{D_{adv}}} \right| = {\displaystyle\frac{1}{4}}\left| {{D_{tar}}} \right|$
  }
  \setlength{\baselineskip}{20pt}
  \item {
  \textit{Adv2:} \small $\left| {{D_{adv}}} \right| = {\displaystyle\frac{1}{2}}\left| {{D_{tar}}} \right|$
  }
  \setlength{\baselineskip}{20pt}
  \item {
  \textit{Adv3:} \small   $   \left| {{D_{adv}}} \right| = {\displaystyle\frac{3}{4}}\left| {{D_{tar}}} \right|$
  }
\end{itemize}

Fig. \ref{fig:adv123} shows the empirical CDF of inference accuracy, precision, and recall of the three different adversaries' membership inference models on the Purchase dataset. It is shown that the more data the adversary has, the higher inference accuracy and precision of the membership inference model against the target model without defense is. Under the protection of the proposed method, the inference accuracy of the three adversaries' membership inference models is between $48\%$ and $54\%$, which is close to the inference accuracy of a random guess. Similarly, the precision and recall of the three adversaries' membership inference models are also greatly reduced. For example, the proposed method significantly reduces the recall of the Adv1's membership inference model from $98.1\%$ to $34.2\%$. Therefore, regardless of the amount of data the adversary has, the proposed method can effectively prevent the adversary from performing membership inference attacks.

\subsection{Defensive Performance against Different Membership Inference Models}\label{sec:exp_diff_MIM}

The proposed method uses the adversarial prediction generated based on the substitute membership inference model to resist the adversary's membership inference models.
To evaluate the effect of similarity between the substitute membership inference model and the adversary's membership inference model on defense performance of the proposed method, and the effect of training data distribution of the substitute membership inference model on defense performance of the proposed method,
we use three different membership inference models (MIM1, MIM2 and MIM3) to evaluate the proposed defense method.
The member data used to train MIM1, MIM2 and MIM3 are the same as the member data used to train MIM0 (which is the adversary's membership inference model used in Section \ref{sec:exp_res}$\sim$Section \ref{sec:exp_diff_data}). The settings of non-member data used to train MIM1, MIM2 and MIM3 are shown in Table \ref{tab:MIMsetting}.
The structure and loss function of the three different membership inference models are also presented in Table \ref{tab:MIMsetting}. Note that, the layer sizes of MIM1 network is [1024, 512, 256, 1]. The other settings of MIM1, MIM2, and MIM3 are the same as that of MIM0.
Table \ref{tab:similar_model} shows the performance of the three membership inference models and the substitute membership inference model.
The recall of MIM1, MIM2 and MIM3 are significantly higher than the recall of the substitute membership inference model. Hence, the performance of MIM1, MIM2 and MIM3 is different from that of the substitute membership inference model.
Next, we evaluate the performance of the proposed method on these different membership inference models.

\begin{table}[htbp]
  \centering
  \footnotesize
  \caption{Detailed Settings of MIM1, MIM2, and MIM3.}
    \begin{tabular}{|m{2.5em}<{\centering}|m{7.5em}<{\centering}|m{6em}<{\centering}|m{8em}<{\centering}|}
    \hline
    \tabincell{c}{Model}   & Structure &  Loss function &  Distribution of non-member data\\
    \hline
    MIM1    &A fully connected
neural network  & Cross-entropy loss function &Same as substitute MIM\\
    \hline
    MIM2    &Same as MIM0& L2 loss function &Same as substitute MIM\\
    \hline
    MIM3      &Same as MIM0& Cross-entropy loss function & Different from substitute MIM\\
    \hline
    \end{tabular}%
  \label{tab:MIMsetting}%
\end{table}%

\begin{table}[htbp]
  \centering
  \footnotesize
  \caption{Performance of the Three Membership Inference Models and the Substitute Membership Inference Model.}
    \begin{tabular}{|c|c|c|c|}
    \hline
    \tabincell{c}{Model}  & \tabincell{c}{Inference \\Accuracy} & Precision & Recall \\
    \hline
    Substitute MIM    & 67.6\% & 69.8\% & 62.2\% \\
    \hline
    MIM1    & 70.4\% & 64.9\% & 88.4\% \\
    \hline
    MIM2    & 75.1\% & 67.7\% & 96.0\% \\
    \hline
    MIM3    & 67.2\% & 60.5\% & 99.3\% \\
    \hline
    \end{tabular}%
  \label{tab:similar_model}%
\end{table}%

\begin{figure*}[htbp]
  \centering
  \includegraphics[width=6.6in]{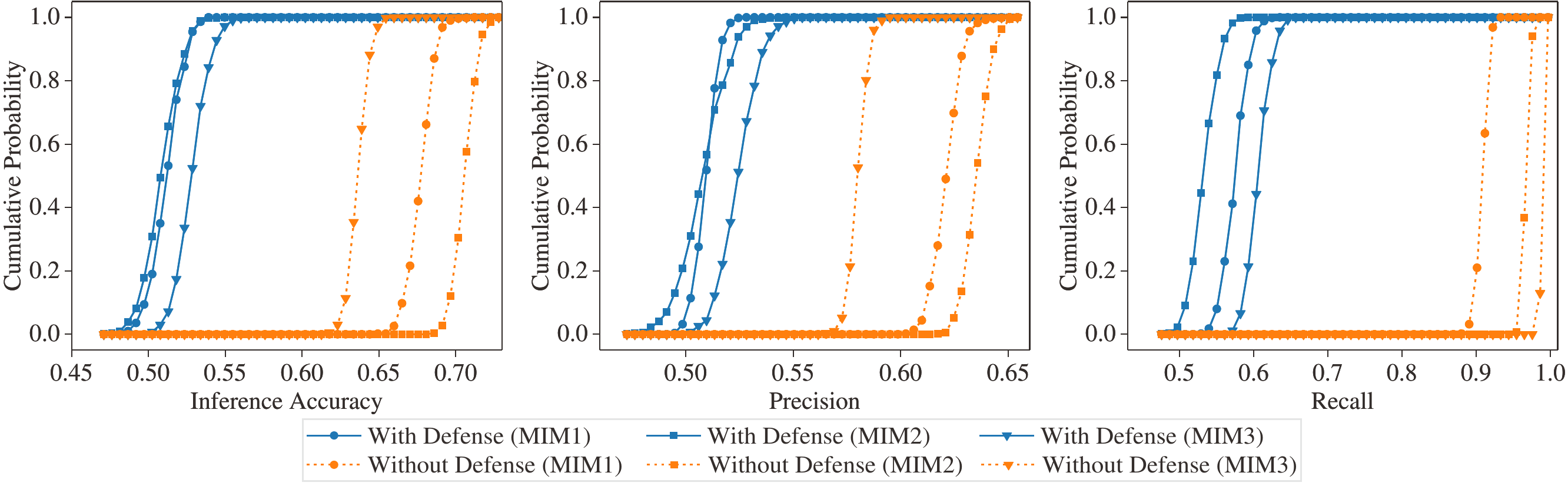}
  \caption{Empirical CDF of inference accuracy, precision, and recall of different membership inference models (MIM1, MIM2, MIM3) on the Purchase dataset.}
  \label{fig:acc_prec_recall_MIM123}
\end{figure*}

Fig. \ref{fig:acc_prec_recall_MIM123} shows the empirical CDF of inference accuracy, precision, and recall of different membership inference models on the Purchase dataset.
MIM1, MIM2 and MIM3 are used to evaluate the effects of the structure of the membership inference model, the loss function of the membership inference model and the distribution of the adversary's non-member data on the defense performance of the proposed method, respectively.
Under the protection of the proposed method, the inference accuracy of MIM1 drops from 67\% to 51\%, and the precision of MIM1 drops from 62\% to 51\%.
The results of MIM1 show that the proposed method can resist against membership inference models with different structures.
This is because the adversarial prediction generated based on the substitute membership inference model has a good transferability, and can mislead different membership inference models.
Under the protection of the proposed method, the inference accuracy of MIM2 drops from 71\% to 51\%, and the precision of MIM2 drops from 64\% to 51\%.
The results of MIM2 show that, even if the loss function used by the adversary's membership inference model is different from that used by the substitute membership inference model, the proposed method can also resist against these membership inference models.
Under the protection of the proposed method, the inference accuracy of MIM3 drops from 64\% to 53\%, and the precision of MIM3 drops from 58.5\% to 53\%.
The results of MIM3 show that the distribution of the adversary's non-member data has a certain impact on the defensive performance of this method. However, the proposed method can still greatly reduce the inference accuracy of MIM3, and resist against MIM3. Therefore, the proposed method has good defensive performance against membership inference models that differ from the substitute membership inference model.

\subsection{Defensive Performance against Adaptive Membership Inference Attacks}\label{sec:adaptive}

Next, we evaluate the performance of the proposed defense method against adaptive membership inference attacks, where the attacker knows the specific defense mechanism. Specifically, we consider the following three adaptive membership inference attacks. Table \ref{tab:adaptive_attacks} shows the performances of the proposed defense under the three adaptive membership inference attacks.

\textbf{Flip attack.} As discussed in Section \ref{sec:APG} (\textit{i.e.}, Equation (\ref{flip_label})), if the substitute membership inference model $I'$ predicts a data $x$ as a member (non-member) of the target model, the proposed AEPPT will add the adversarial perturbations $\delta_{t}$ on the prediction.
In this way, the generated adversarial perturbation will mislead the adversary's membership inference model to infer the data $x$ as a non-member (member) of the target model.
However, an adaptive adversary can simply flip the prediction result of his membership inference model to launch the inference attacks, \textit{i.e.}, the data $x$ is considered to be a non-member if the membership inference model predicts it as a member.

However, such adaptive membership inference attack will not work under the defense of our proposed method. The reason is that, as mentioned in Section \ref{sec:exp_res}, the proposed AEPPT can reduce the inference accuracy of membership inference model to the random guess (52\% on CIFAR100 dataset and 51\% on Purchase dataset).
Since all the prediction results of adversary's membership inference model will be flipped, the inference accuracy of such adaptive attack will be flipped as well.
Specifically, in our experiments, the accuracy of flip attack on CIFAR100 and Purchase datasets is 48\% (100\%-52\%) and 49\% (100\%-51\%), respectively.
In this way, even the attacker knows the defense mechanism and flips the prediction results, the accuracy of the flip attack is still close to that of a random guess (\textit{i.e.}, 50\%).

\begin{table}[htbp]
  \centering
  \footnotesize
  \caption{Performances of the Proposed AEPPT under Three Adaptive Membership Inference Attacks.}
    \begin{tabular}{|c|c|c|c|c|}
    \hline
    \multicolumn{2}{|c|}{Adaptive Attacks} & \tabincell{c}{Flip \\Attack} & \tabincell{c}{Rounding \\Attack} & \tabincell{c}{Adversarial \\Training Attack} \\
    \hline
    \multicolumn{1}{|c|}{\multirow{2}[2]{*}{\tabincell{c}{Inference \\Accuracy}}} & CIFAR100 & 48\%  & 49.71\% & 50.79\% \\
\cline{2-5}          & Purchase & 49\%  & 40.46\% & 62.90\% \\
    \hline
    \end{tabular}%
  \label{tab:adaptive_attacks}%
\end{table}%

\textbf{Rounding attack \cite{shokri2017membership, jia2019memguard}.}
The proposed method adds the perturbation to each component of an original prediction to generate the adversarial prediction, so as to mislead the adversary's inference model.
The adaptive attacker can round the generated adversarial prediction to remove these added perturbations, and then use these rounding predictions to train his membership inference model.
Specifically, in our experiments, we assume that the attacker rounds each confidence score in the prediction to 3 decimals.

The experimental results in Table \ref{tab:adaptive_attacks} show that, the inference accuracy of rounding attack is as low as 49.71\% (CIFAR100) and 40.46\% (Purchase), respectively.
This indicates that even the adaptive attacker truncates the generated adversarial prediction, the proposed AEPPT method is also effective, and can significantly degrade the inference accuracy of membership inference attacks.

\textbf{Adversarial training attack \cite{Athalye2018Obfuscated}.} The adversarial training is considered to be an effective approach against the adversarial examples attacks \cite{Athalye2018Obfuscated}.
Therefore, an adaptive attacker can exploit the proposed adversarial prediction generation algorithm (\textit{i.e.}, Algorithm 1 in Section \ref{sec:APG}) to construct the adversarial perturbations.
Then, those generated adversarial predictions can be used to train his membership inference model to make it robust.
As shown in Table \ref{tab:adaptive_attacks}, under this complex attack scenario, the inference accuracy of adversarial training attack on CIFAR100 and Purchase datasets is only 50.79\% and 62.90\%, respectively, which indicates that such adaptive attack is failed.

Overall, under these more stronger adaptive membership inference attacks, where the adversary knows the defense mechanism and attempts to eliminate the effectiveness of the proposed method, the proposed AEPPT is demonstrated to be robust and can greatly reduce the inference accuracy of these adaptive attacks.

\subsection{Comparison with Other Defense Methods}\label{sec:discussion}

\begin{figure*}[htbp]
  \centering
  \subfigure[CIFAR100 Dataset]{
    \label{}
    \includegraphics[width=2.45in]{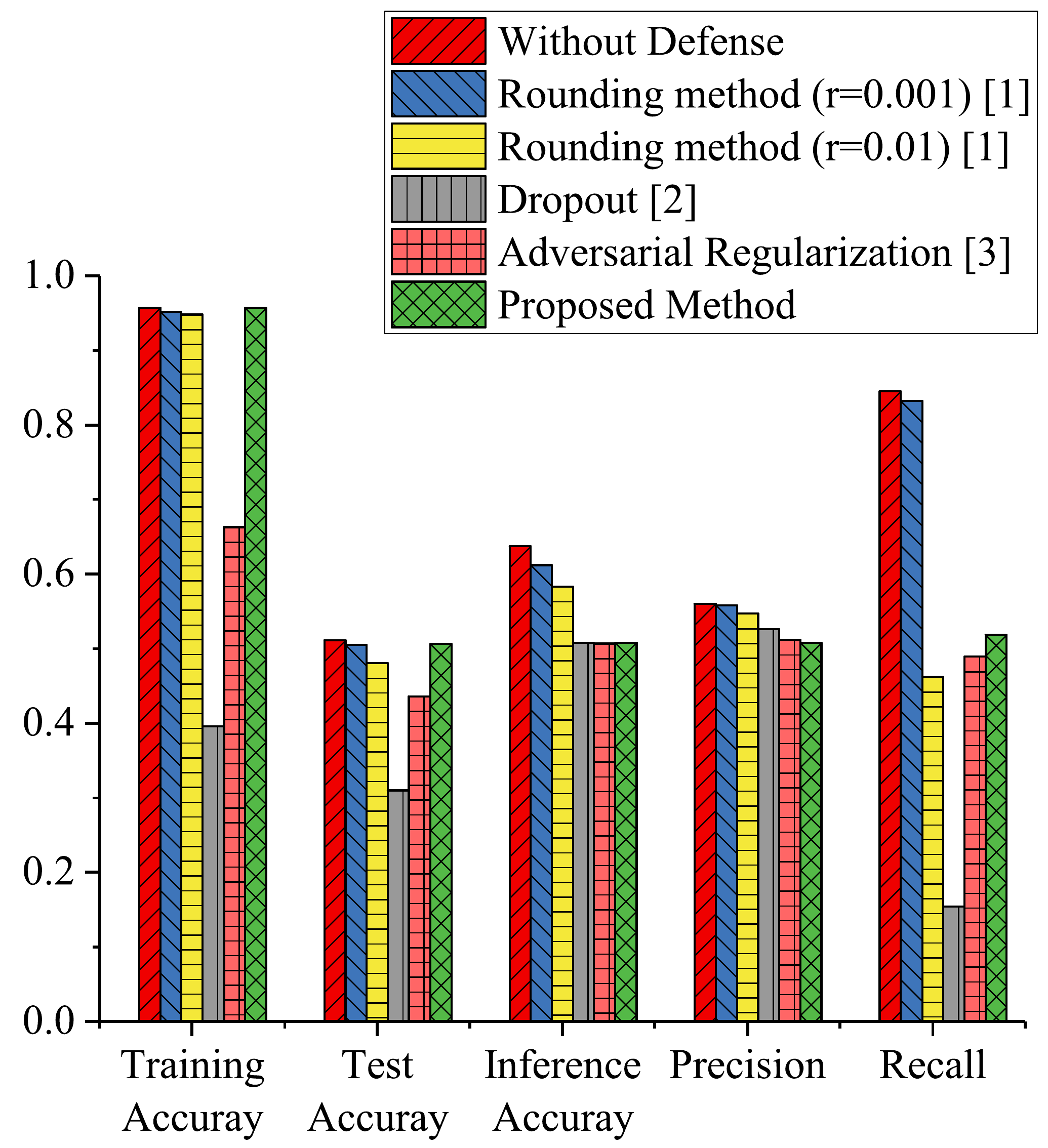}}
    \hspace{0.3in}
  \subfigure[Purchase Dataset]{
    \label{}
    \includegraphics[width=2.45in]{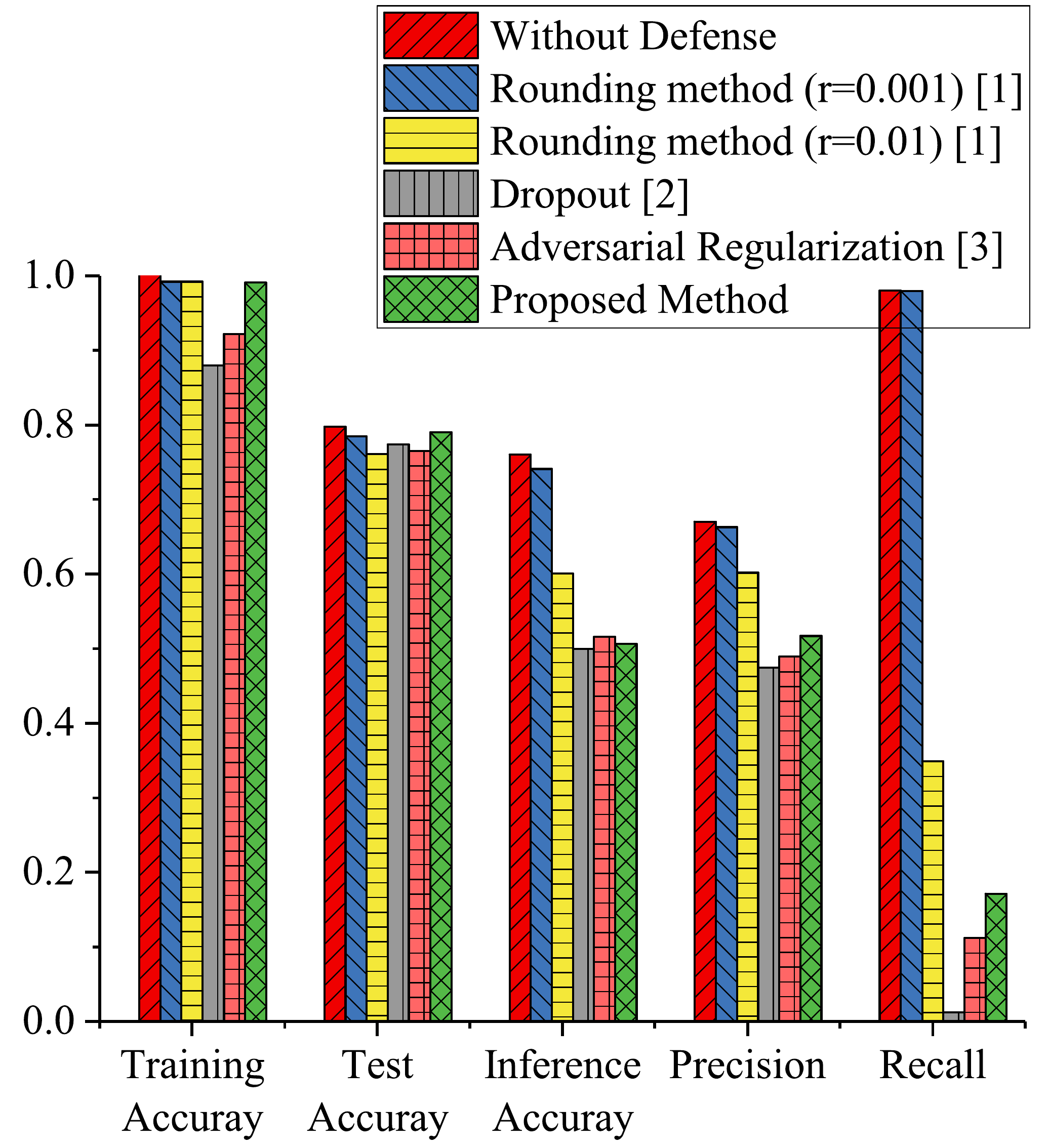}}
  \caption{Comparison between the proposed defense method and the state-of-the-art defense methods (\cite{shokri2017membership}, \cite{salem2018ml}, and \cite{nasr2018machine}) from the following aspects: the training accuracy and test accuracy of the target model, and the inference accuracy, precision and recall of the membership inference model.}
  \label{fig:comp}
\end{figure*}

Finally, we compare the proposed method with the state-of-the-art defense methods: 1) methods of modifying the output of the target model; 2) methods of preventing the target model from overfitting. Particularly, the method of modifying the output of the target model is the rounding method (\textit{i.e.}, truncating method) \cite{shokri2017membership}. We round the confidence scores in the prediction to 3 floating points (\textit{i.e.}, $r=0.001$) and 2 floating points (\textit{i.e.}, $r=0.01$), respectively. The methods of preventing the target model from overfitting are the dropout method \cite{salem2018ml} and the adversarial regularization method \cite{nasr2018machine}.
To compare the experimental results, in the experiment, we adopt the same experimental setup as in \cite{nasr2018machine}.
For the CIFAR100 dataset, the number of the adversary's data is half of the training data of the target model (\textit{i.e.}, {\small $\left| {{D_{adv}}} \right| = {1 \over 2}\left| {{D_{tar}}} \right|$}), and the structure of the target model is AlexNet \cite{krizhevsky2012imagenet}.
For the Purchase dataset, the number of the adversary's data is one quarter of the training data of the target model (\textit{i.e.}, {\small $\left| {{D_{adv}}} \right| = {1 \over 4}\left| {{D_{tar}}} \right|$}), and the structure of the target model is a 4-layer fully connected neural network \cite{nasr2018machine}. The membership inference model is MIM0 (as described in Section \ref{sec:MIM_Intro}).
In our experiments, the data used to evaluate the membership inference attacks under those defenses are randomly selected from the training data $D_{tar}$ and non-training data $\bar{D}_{tar}$ of the target model.
Specifically, we select 10,000 member data and 5,000 non-member data from CIFAR100 dataset, and select 10,000 member data and 10,000 non-member data from Purchase dataset, respectively.

Fig. \ref{fig:comp} shows the comparison between the proposed defense method and other defense methods (the rounding method \cite{shokri2017membership}, the dropout method \cite{salem2018ml}, and the adversarial regularization method \cite{nasr2018machine}).
For the rounding method with $r=0.001$, the method cannot reduce the inference accuracy, precision and recall of the membership inference model. In other words, the rounding method with $r=0.001$ cannot resist against membership inference attacks. For the rounding method with $r=0.01$, the method slightly reduces the performance of the membership inference model. The inference accuracy of the membership inference model drops from 63.7\% to 58.3\% on the CIFAR100 dataset, and drops from 76.0\% to 60.0\% on the Purchase dataset. Therefore, the rounding method can slightly mitigate membership inference attacks only when the number of floating points in confidence score is set to be less than or equal to 2, but it cannot effectively resist membership inference attacks.
Moreover, the rounding method makes the model's predictions inaccurate. Specifically, when the floating point number in confidence score is less than or equal to 2, the user can only obtain an approximate confidence returned by the target model. In this way, the rounding method significantly reduces the utility of the model's predictions and the quality of the MLaaS.
Compared with the rounding method \cite{shokri2017membership}, the proposed method can significantly reduce the inference accuracy and precision of the membership inference model to around 50\% (\textit{i.e.}, like a random guess), which means the membership inference attacks completely failed.
Besides, the perturbation in the adversarial prediction generated by the proposed method is extremely small. The adversarial prediction is close to the original prediction of the target model, thus doesn't affect the quality of the MLaaS.
In terms of the training accuracy and test accuracy of the target model, the rounding method \cite{shokri2017membership} and the proposed method hardly affect the accuracy of the target model.
In conclusion, compared with the rounding method \cite{shokri2017membership}, the defensive performance of the proposed method is much better than that of the rounding method \cite{shokri2017membership} and doesn't affect the performance or quality of the model.

For the dropout method \cite{salem2018ml} and the adversarial regularization method \cite{nasr2018machine}, the defensive performance of these two methods \cite{salem2018ml,nasr2018machine} is close to that of the proposed method. These methods can also reduce the inference accuracy and precision of the membership inference model to about $50\%$ on CIFAR100 and Purchase datasets. However, the dropout method \cite{salem2018ml} and the adversarial regularization method \cite{nasr2018machine} affect the accuracy of the target model on training data and test data.
Compared with these two methods \cite{salem2018ml,nasr2018machine}, the proposed method doesn't affect the training accuracy and test accuracy of the target model. The reason is as follows. To ensure that the prediction of the target model is indistinguishable, the dropout method \cite{salem2018ml} and the adversarial regularization method \cite{nasr2018machine} need to make the training accuracy of the target model close to the test accuracy, which will sacrifice the performance of the target model on the training data and test data. For the proposed method, it makes the generated adversarial prediction indistinguishable by adding small perturbations to the original prediction of the target model. The added perturbations only mislead the membership inference model, but do not affect the accuracy of the target model.
In summary, compared with other defense methods, the proposed method can effectively protect the privacy of the target model's training data without affecting the performance and utility of the target model.

Note that, there are several insurmountable difficulties when comparing the proposed AEPPT with the MemGuard method \cite{jia2019memguard}.
First, the work \cite{jia2019memguard} does not provide the detailed parameter settings of their optimization function, and the experiments of the two works are deployed on different platforms (Tensorflow for \cite{jia2019memguard} and Pytorch for this paper), thus we cannot reproduce their whole implementations in our local devices.
Second, the target model and experimental datasets of work \cite{jia2019memguard} are different from ours, which makes it difficult to compare the MemGuard method \cite{jia2019memguard} with the proposed AEPPT under the same evaluation conditions.
Nonetheless, we have compared the concurrent work \cite{jia2019memguard} with this paper in three different aspects, and the comparison results are concluded in Table \ref{tab:compare_current}.

\begin{table}[htbp]
  \centering
  \footnotesize
  \caption{Comparison of the Proposed AEPPT Method and the MemGuard Method \cite{jia2019memguard}.}
    \scalebox{0.93}{
    \begin{tabular}{|c|c|c|c|}
    \hline
    \multicolumn{2}{|c|}{Defenses} & MemGuard \cite{jia2019memguard} & \multicolumn{1}{c|}{AEPPT} \\
    \hline
    \multicolumn{2}{|c|}{\tabincell{c}{Defensive performance}} & Around 50\% & \multicolumn{1}{c|}{51\%} \\
    \hline
    \multicolumn{2}{|c|}{\tabincell{c}{Size of added perturbations}} & 0.8   & \multicolumn{1}{c|}{0.2637} \\
    \hline
    \multicolumn{1}{|c|}{\multirow{2}[7]{*}{\tabincell{c}{Adversary's \\knowledge}}} & Training data & 20\%  & \multicolumn{1}{c|}{50\%} \\
\cline{2-4}  & Target model & Black-box & \tabincell{c}{Model's structure \& \\training algorithm} \\
    \hline
    \end{tabular}}%
  \label{tab:compare_current}%
\end{table}%

First, the MemGuard \cite{jia2019memguard} and the proposed AEPPT can achieve the similar defensive performance. Both of which can degrade the inference accuracy of membership inference attacks to that of a random guess, \textit{i.e.}, 50\%.
Second, the defensive performance of MemGuard \cite{jia2019memguard} improves as the size of added perturbations (also called confidence score distortion in \cite{jia2019memguard}) increases.
More specifically, to reduce the accuracy of adversary's membership inference attacks to around 50\%, the $L_1$-norm of added perturbations in their three experimental datasets all increases to 0.8 \cite{jia2019memguard}.
However, for the proposed AEPPT method, the size of generated adversarial perturbations is only as low as 0.2637 on CIFAR100 dataset.
This indicates that the proposed AEPPT method only needs to add smaller modifications on the original prediction of the target model than MemGuard method \cite{jia2019memguard}), which makes the added perturbations more difficult to be noticed by the adversary.
Finally, the work \cite{jia2019memguard} considers the adversary only has black-box access to the target model, while this paper assumes a more powerful attacker that knows the target model's structure and the training algorithm.
Moreover, compared to the work \cite{jia2019memguard} where the adversary is assumed to know only 20\% of the training data, this paper considers a stronger inference attacker who knows half of the target model's training data (\textit{i.e.}, {\small $\left| {D}_{adv} \right| =  \frac{1}{2} \left| {D}_{tar} \right|$}).
Therefore, the proposed AEPPT method can defeat more powerful attackers.

\section{Conclusion}\label{sec:conclusion}
We propose a novel and effective method to preserve the privacy of model's training data resisting membership inference attacks. Inspired by the adversarial example attack which was used to be an attack method, we present an adversarial prediction generation algorithm for privacy protection of the machine learning models.
The proposed method is general, and does not modify the architecture or the training process of the target model.
Experiment results on CIFAR100 and Purchase datasets show that the proposed method can make the inference accuracy and precision of the adversary's membership inference model close to that of a random guess, and also greatly reduce the recall of the membership inference model. Evaluation under different experimental settings indicates that the number of the target model's output classes, the number of the adversary's data, the training data distribution, and different adversary's membership inference models hardly affect the performance of the proposed defense method.
\textit{blue}{Moreover, the proposed AEPPT is also demonstrated to be effective under three adaptive attacks where the adversary knows the defense mechanism.}
In summary, the proposed defense method can effectively resist membership inference attacks without affecting the normal performance of the model or the quality of the model's services.
Privacy-related attacks on machine learning models pose serious threats to both the model provider and the user. Privacy-preserving machine learning techniques in various scenarios, including distributed or Internet of Things scenarios, are important and open topics.
In future works, we will study defense techniques against other privacy related attacks on machine learning models.

\bibliographystyle{IEEEtran}
\bibliography{ref}

\end{document}